\documentclass[sn-mathphys,Numbered]{sn-jnl}

\usepackage{graphicx}%
\usepackage{multirow}%
\usepackage{amsmath,amssymb,amsfonts}%
\usepackage{amsthm}%
\usepackage{mathrsfs}%
\usepackage[title]{appendix}%
\usepackage{xcolor}%
\usepackage{textcomp}%
\usepackage{manyfoot}%
\usepackage{booktabs}%
\usepackage{algorithm}%
\usepackage{algorithmicx}%
\usepackage{algpseudocode}%
\usepackage{listings}%

\usepackage{hyperref}

\begin{document}

\title{The Role of Electric Grid Research in Addressing Climate Change}

\author*[1,2]{\fnm{Le} \sur{Xie}}\email{xie@seas.harvard.edu}
\author[1]{\fnm{Subir} \sur{Majumder}}\email{subir.majumder@tamu.edu}
\author[3]{\fnm{Tong} \sur{Huang}}\email{thuang7@sdsu.edu}
\author[1]{\fnm{Qian} \sur{Zhang}}\email{zhangqianleo@tamu.edu}
\author[4]{\fnm{Ping} \sur{Chang}}\email{ping@tamu.edu}
\author[5]{\fnm{David} \sur{J. Hill}}\email{davidj.hill@monash.edu}
\author[6]{\fnm{Mohammad} \sur{Shahidehpour}}\email{ms@iit.edu}

\affil[1]{\orgdiv{Department of Electrical and Computer Engineering}, \orgname{Texas A\&M University}, \orgaddress{\city{College Station}, \postcode{77843}, \state{Texas}, \country{United States}}}
\affil[2]{\orgdiv{Harvard John A. Paulson School of Engineering and Applied Sciences}, \orgname{Harvard University}, \orgaddress{\city{Cambridge}, \postcode{02134}, \state{Massachusetts}, \country{United States}}}
\affil[3]{\orgdiv{Department of Electrical and Computer Engineering}, \orgname{San Diego State University}, \orgaddress{\city{San Diego}, \postcode{92182}, \state{California}, \country{United States}}}
\affil[4]{\orgdiv{Department of Oceanography}, \orgname{Texas A\&M University}, \orgaddress{\city{College Station}, \postcode{77843}, \state{Texas}, \country{United States}}}
\affil[5]{\orgdiv{Department of Electrical and Computer Systems Engineering}, \orgname{Monash University}, \orgaddress{\city{Clayton}, \postcode{3800}, \state{Victoria}, \country{Australia}}}
\affil[6]{\orgdiv{Robert W. Galvin Center for Electricity Innovation}, \orgname{Illinois Institute of Technology}, \orgaddress{\city{Chicago}, \postcode{60610}, \state{Illinois}, \country{United States}}}


\abstract{

Addressing the urgency of climate change necessitates a coordinated and inclusive effort from all relevant stakeholders. Critical to this effort is the modeling, analysis, control, and integration of technological innovations within the electric energy system, which plays a crucial role in scaling up climate change solutions. This perspective article presents a set of research challenges and opportunities in the area of electric power systems that would be crucial in accelerating Gigaton-level decarbonization. Furthermore, it highlights institutional challenges associated with developing market mechanisms and regulatory architectures, ensuring that incentives are aligned for stakeholders to effectively implement the technological solutions on a large scale.

}

\maketitle

\section*{Main}
The electricity sector plays two pivotal roles in tackling climate change. First, according to the recent IPCC report \cite{SourcesGreenhouseGas}, as of 2019, the electricity and
heating sector contributes to approximately 23\% of global greenhouse gas emissions. Therefore, cleaning up the electricity sector itself is a major step towards the goal of reducing gigaton-level carbon emissions for the entire planet by 2050. Second, for many carbon management or reduction technologies, the best route to achieve a speedy and scalable impact is through large-scale integration into the electric grid.

For the first role, significant decarbonization efforts are ongoing to replace fossil fuel-based generation technologies with renewable energy resources such as wind and solar \cite{gielen2019role, aemo2024_ISP}. Efforts are also underway to incorporate carbon management technologies such as point source carbon capture, carbon transport and storage, carbon dioxide removal and conversion, and hydrogen \cite{doe}. For example, the U.S. Department of Energy (DOE)  developed goals to achieve more than 95\% carbon capture at sources of carbon emissions at power plants \cite{doe2021}. For the second role, the electricity sector plays another increasingly important role in supporting more electrification of energy demand that comes from transportation \cite{wei2021personal,xie2021toward}, industrial heating/cooling \cite{amonkar2023differential}, computing industries \cite{iea}, and many household appliances. Much of these efforts will need to be scaled up and integrated with the electric grid infrastructure.

\begin{figure}[H] 
\centering
  \includegraphics[scale=0.44]{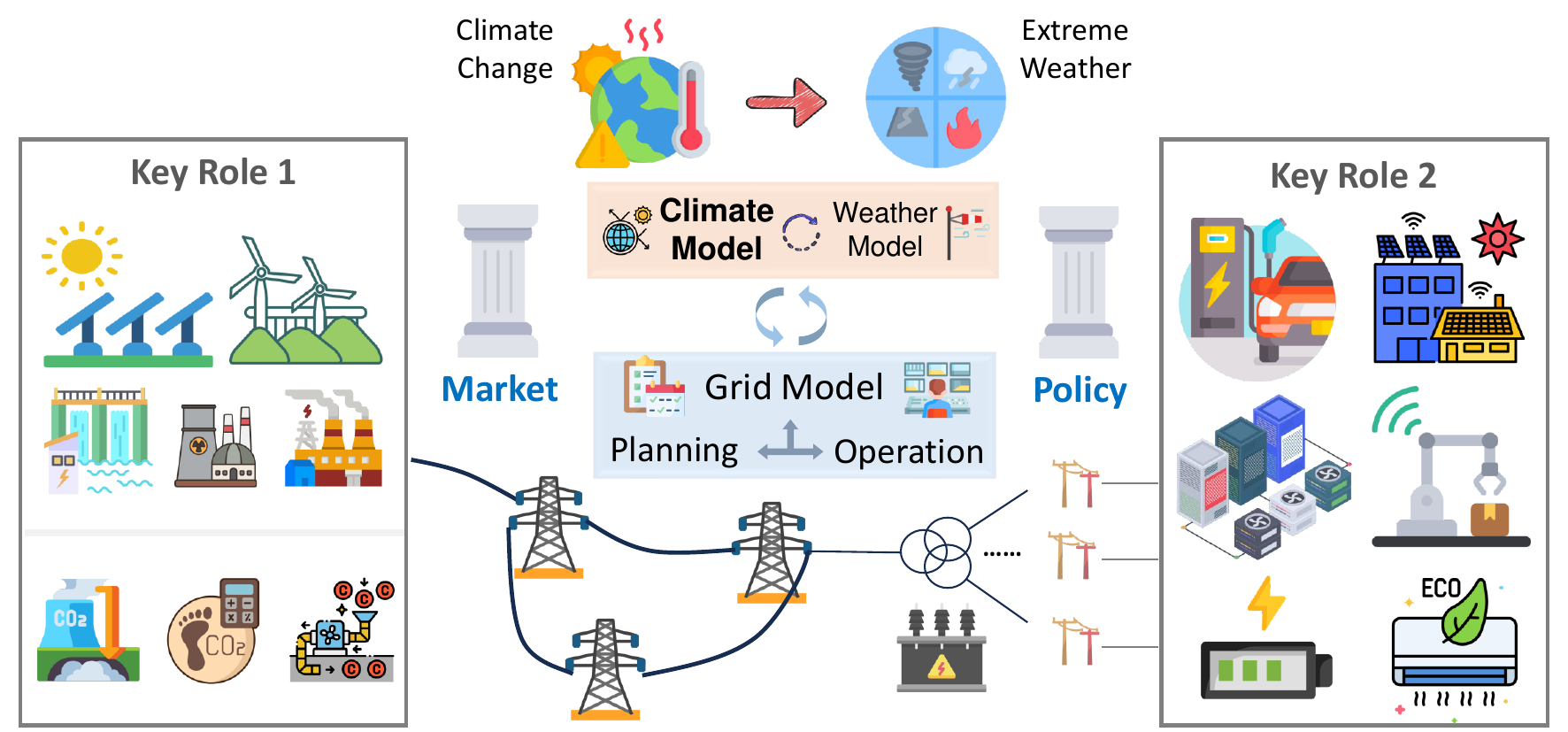}
  \caption{\textbf{The crucial role of the electric grid research in tackling climate change.} This schematic illustrates how electric grid research interacts with climate change research. `Key Role 1' represents the decarbonization of the power generation sector, while `Key Role 2' denotes the integration of other sectors—such as transportation, industrial heating/cooling, computing loads, and household appliances—into the electric grid, facilitating their decarbonization. We show that climate and weather models must be utilized for power grid operations and planning, with the power grid providing essential feedback to enhance the climate models. Finally, a climate-aware market redesign and policy support are essential to leveraging the power grid in combating climate change. The icons used in this figure are sourced from \url{flaticon.com}.}
  \label{timeslot}
\end{figure}

Therefore, achieving gigaton-level carbon emission reduction through the power grid necessitates the expansion of the electrical systems, which is now more intertwined with weather and climate than ever before (Figure \ref{timeslot}). For example, due to their variable nature of renewable resources, integrating them into the power grid significantly impacts planning processes, which used to operate under the generation-following-demand paradigm \cite{osti_1972011}. The emergence of long-term energy storage \cite{energy.gov_2023} and demand response technologies \cite{9208672,leeTargetedDemandResponse2022,yilmaz2022analysing} has led to the concept of demand-following-generation as an added mechanism to achieve overall power balancing during operation. However, the necessary long-term storage for the power grid, e.g., for `dunkelflaute’ events \cite{hutson2022renewable}, can only be accurately planned through climate model simulations \cite{li2021mesoscale}. In this article, we ask two key questions. First, what power grid researchers should prioritize as we integrate more and more carbon-neutral technologies? Second, how do we effectively integrate climate research with power grid research to address the emerging complexities? The following two sections provide an answer to these two questions.


\section*{Electric grid research challenges to address climate change}

There are three key power grid research challenges in addressing climate change. First, power grid researchers still use outdated weather patterns to generate scenarios for planning, while climate change research has demonstrated that weather patterns are changing and impacting renewable generation and power system demand. Second, the overall operational performance of the grid is not well understood when integrating renewables or electrifying transportation and heating and cooling systems, especially with a wide array of heterogeneous inverter-based resources such as solar and wind. Third, climate change research typically focuses on the long term, whereas power system research tends to be more short-term. Unless the incentives are properly aligned through markets and policy designs, it would be extremely difficult to collectively address climate change issues.

\subsection*{Lack of tailored climate simulation for long-term planning}\label{sec21}

Addressing climate change requires renewable energy sourced from areas far from load centers to satisfy ever-growing load demands, necessitating the expansion of transmission line capacity and the building of more renewable energy harvesting farms. Many \cite{building_better_grid_initiative, ARENA2023} of the world's governments are actively incentivizing these activities.

Power system expansion planning is a well-researched \cite{9238453} area, but these planning activities utilize widely used weather patterns for scenario generation, and these weather patterns can indeed change due to climate change. For example, future weather statistics could include increased ``energy drought'' occurrence frequencies (e.g., `dunkelflaute’ events \cite{hutson2022renewable}), posing significant risks to energy security \cite{brownMeteorologyClimatologyHistorical2021}. With the expected notable ``west-to-east interhemispheric shift'' in mean wind power potential and an overall increase in solar energy potential \cite{leiCobenefitsCarbonNeutrality2023}, transmission lines built without consideration of these climate dynamics would most likely remain underutilized. Climate change can lead the typical summer-peaking states into dual or winter-peaking states \cite{keskar2023planning}. Increasing frequency of extreme weather events in certain regions, like the Texas grid outage during winter storm Uri, underscores the necessity of substantial backup power capacity to ensure future grid reliability \cite{xu2024resilience} and resiliency \cite{stankovic2022methods}. We also need to utilize multimodal climate change models to capture changes in consumer demands for resource expansion planning \cite{cohen2022multi}. Therefore, system planning encompassing the expansion and retirement of generation units, transmission, storage, demand management, and other new technologies should incorporate reliable climate and weather predictions.

Fortunately, there have been rapid advancements in climate models, and computational capabilities are progressing rapidly. A new report \cite{levin2023extreme} to the President from the U.S. President’s Council of Advisors on Science and Technology (PCAST) highlights the considerable potential for improving predictions of the likelihood of extreme weather events using high-resolution climate models. Disseminating this information to households, businesses, and government agencies would help grid operators better manage the electric grid, even with limited resources. While existing electric grid studies, such as the US-DOE report on National Transmission Needs \cite{GDO_TransmissionNeedsStudy}, also advocate for enhanced infrastructure, the key research challenge lies in incorporating climate-related factors into long-term planning strategies.

\subsection*{Lack of system-aware, grid-edge operations of inverter-based resources}

Many clean energy resources, such as solar panels, wind turbines, and battery storage, require power electronics inverters to interface with power grids. As a result, these inverter-interfaced clean energy resources, known as inverter-based resources (IBRs) \cite{9762253}, are key technology enablers of electricity infrastructure decarbonization. Residential IBRs are typically connected to medium/low-voltage distribution systems. Examples include rooftop solar panels, batteries rated in kilowatts, electric vehicles (EVs), and their charging infrastructures. During extreme weather events that disrupt bulk electricity infrastructure, residential IBRs are expected to self-organize to establish small-scale grids, such as microgrids \cite{farrokhabadi2019microgrid} to ensure continued power supply to end-users. Residential IBRs empower consumers by turning them into producers, democratizing energy production and reducing transmission losses by generating power close to where it is consumed. However, as more solar capacity has come online, grid operators have observed a drop in net load due to the generation from residential solar panels, when power generation from utility-scale solar farms tends to be highest. Such an imbalance between generation and load leads to the famous California duck curve \cite{caiso_2016_fast_facts}, which can compromise energy security and economic efficiency. This imbalance is also evident in the increasing frequency of negative net load and associated negative prices \cite{AEMO2024_QED} in the electricity market of Australia. Clearly, a lack of system-aware control has the potential to exacerbate power network operations.

As major coal-fired plants are scheduled to retire almost every year over the next decade, the existing location and electrical infrastructure could be utilized by the utility-scale IBRs, such as renewable energy zones (REZs) \cite{simshauser2021renewable}. REZs indeed solve some of the drawbacks of residential IBRs. However, the control systems of today’s commercial IBRs are tuned by manufacturers by overly simplifying the dynamics of the host system. As a result, when networking with other IBRs, the locally well-tuned IBRs may conflict with their peers \cite{mohammadpour2015analysis}. The key controller algorithms are almost always unavailable to protect their intellectual property \cite{kroposki2022gridforming}. Therefore, the key research question is how to avoid system-level issues when multiple proprietary IBRs are networked in a grid. This problem is especially exacerbated when thousands of ``behind the meter'' IBRs try to coordinate with each other. Centralized control by utilities is impractical in this case \cite{navidi2023coordinating}. Designing system-aware controls for IBRs, with minimal information exchange, becomes crucial to prevent system-level issues at the device-design stage rather than relying on post-event mitigations.

\subsection*{Temporal misalignment of market and policy design in grid and climate systems}

In the realm of the electricity industry, market and policy decision-making typically operate on a time scale ranging from days to years, reflecting the immediate and intermediate needs of grid management and energy distribution. In stark contrast, the design of market strategies and policies for addressing climate change encompasses a far more extended timeline, often spanning several decades. This discrepancy creates a significant challenge in synchronizing the short-term operational strategies of the electricity sector with the long-term objectives of climate policy \cite{joskow2022hierarchies,li2023energy}.

First, aggressive long-term climate targets may conflict with short-term stakeholders' rights in the power grid. For example, in Australia, the market operator AEMO, through their integrated system plan (ISP) \cite{aemo2024_ISP}, shows that coal-based power plants are to be withdrawn by 2038 only to be replaced by grid-scale wind and solar, rooftop solar photovoltaics, and storage. However, the rapid uptake of grid-scale wind and solar is not enforceable due to the vertically disintegrated nature of the Australian power market. While the state governments in Australia are urgently forging ahead with Renewable Energy Zones (REZs) to replace retiring coal plants with renewable-based farms and build more transmission lines \cite{ARENA2023}, there have been pushbacks from certain communities about disproportionately burdening renewable-rich areas with aerial aesthetic displeasure \cite{TheAge2023}.

Second, climate change introduces greater uncertainty into power system planning, and the power markets are ill-equipped to guide generation investment, which may not ensure resource adequacy in the future. In energy-only markets, such as the Texas electricity market, there has been recent discussion about implementing a `performance credit mechanism' to ensure long-term grid reliability, but the initial draft did not incorporate impacts of extreme weather events such as Winter Storm Uri \cite{walton_texas_2022}. On the other hand, introducing a capacity market might secure adequate capacity to meet desired reliability criteria, such as the loss of load expectation (LOLE) of one day in ten years \cite{joskow2006competitive}. However, these criteria have two major issues: (i) they are relatively short-term compared to climate policies, and (ii) they seem to ignore low-probability events, such as extreme weather events. Furthermore, the efficiency and fairness of capacity markets heavily depend on the accreditation of different types of generation technologies \cite{isone} and the modeling of the demand curve \cite{zhao2017constructing}, which are extremely complex tasks.

Bridging the temporal gap in power grid decarbonization and emission-related policies is extremely crucial for aligning immediate energy needs with enduring environmental targets, thereby enhancing the effectiveness of interventions in both domains. Without such integration, efforts in either area risk being undermined by conflicting priorities and uncoordinated policies, potentially stalling progress on both power system security and climate change mitigation.

\section*{Interaction between power system and climate researchers}

Power system researchers can interact and collaborate with climate researchers in three unique ways. First, power system researchers should utilize climate data for planning. Second, climate researchers can incorporate power grid operational data to refine their climate models, and power system researchers can assist climate researchers in identifying critical areas to focus on in their climate models. Third, by utilizing climate simulation data, power system researchers and economists can develop innovative products and mechanisms tailored for short-term efficient and reliable energy system operation while meeting long-term environmental goals.

\subsection*{Climate-informed planning for enhanced resiliency and reliability } 

First, power grids around the world evolved to cater to regional energy demand, supported by local policy mandates and unique generation characteristics (e.g., the Pacific Northwest in the United States has plenty of hydro-electric potential \cite{eia_washington}, western Texas has an abundance of wind resources \cite{eia_texas}). Second, global renewable energy potential is not uniform, and the distribution can be impacted by climate change \cite{leiCobenefitsCarbonNeutrality2023}. Third, while climate change is a planetary-scale problem, each region faces its unique climate-related challenges (e.g., the wildfires in California \cite{california_energy_commission_2019} and hurricanes on the Gulf and east coast of the United States \cite{department_of_energy_2013}). Therefore, the interdependence between energy and climate systems can result in regional energy security concerns with the growing prominence of renewable energy sources, and power system planners need to be cognizant of this relationship while planning.

\begin{figure*}[hbt!]
    \centering
    \includegraphics[width = 0.95\textwidth]{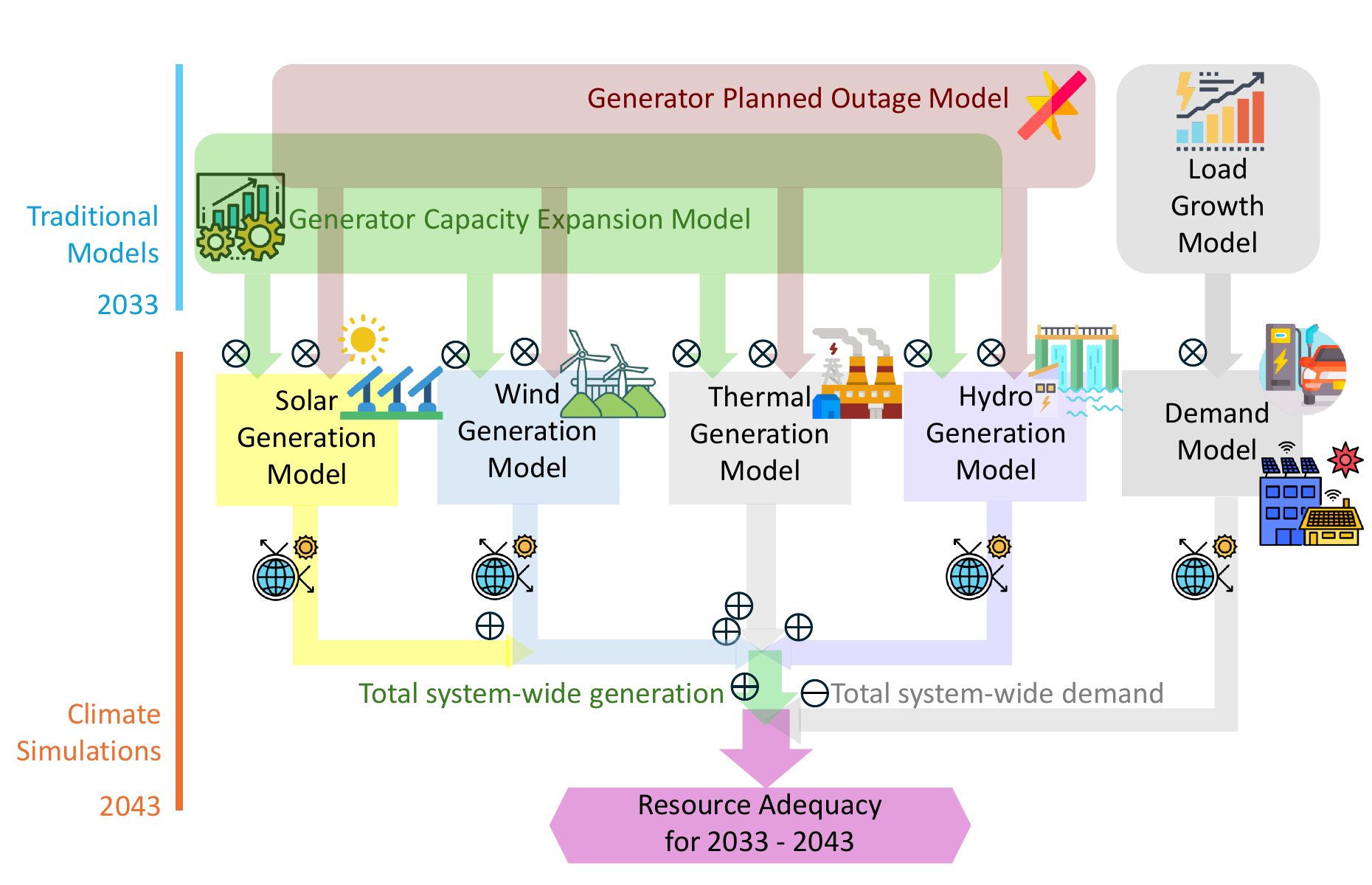}
    \caption{ \textbf{Power grid resource adequacy assessment utilizing climate data.} The climate data, obtained from both high-resolution (0.25 degrees) and low-resolution (1 degree) global climate simulations, include atmospheric temperature, relative humidity, dew point, wind speed, and solar radiation flux. These datasets enable more accurate predictions of renewable generation and load demands for the power grid, e.g., from 2033 to 2043. They enhance traditional resource adequacy frameworks that already include load growth models, generator capacity expansion models, and generator planned outage models, which are essential for determining effective system-wide loads and generations. By integrating these comprehensive climate datasets with traditional models, we can better assess resource adequacy and identify potential reliability issues under various climate scenarios. Figure adapted from ref. \cite{zheng2023impact}. The icons used in this figure are sourced from \url{flaticon.com}. }
    \label{fig:diagram}
\end{figure*}

In regards to utilizing climate data, a recent study \cite{zheng2023impact} provides a framework for power system resource adequacy analysis (Figure \ref{fig:diagram}). This study identified that while low-resolution climate simulations can indeed provide a rough estimate of system reliability, high-resolution simulations can provide a more informative assessment of low-probability, high-impact extreme events. However, both high and low-resolution assessments suggest the need to prepare for severe blackout events in winter due to extremely low temperatures. Changing weather patterns due to climate change, as discussed earlier, could be similarly incorporated in the power system planning studies. Another recent research explores multiple scenarios based on climate models, revealing that the current placement of renewable energy farms in Australia is suboptimal \cite{gunn2023spatial}. These preliminary findings demonstrate the importance of incorporating higher-resolution climate simulations for a reliable and robust climate-informed analysis for resiliency and reliability in power system planning.

\vspace{0 pt}

Future research opportunities to understand the impacts of climate change on the power grid include (a) quantifying the reliability and resilience of specific long-term planning schemes, along with investigating statistical confidences and the effectiveness of various ways to represent uncertainty, (b) identifying the most critical potential climate conditions for subsequent measures to enhance accordingly, and (c) performing sensitivity analyses of the resulting reliability and resiliency indices concerning the uncertainty of climate projections. Based on the outcomes of these activities, a unified planning approach can be developed that accounts for climate change risks and provides a solid foundation for key decisions in planning. These key insights include the minimum required energy storage and demand response programs for policymakers, system operators, and market participants. While the approaches to achieving a zero-carbon solution vary, simply a power grid tailored to local weather (operations) and climate (planning) ensures greater renewable electrification, which will all aggregate globally to arrest climate change. Additionally, digitization and machine learning techniques will enhance the scalability and expedite the development of solutions within both climate and power research domains. This includes accelerating computationally demanding simulations of hybrid climate and electricity models \cite{rolnick2022tackling} and generating synthetic open-source data  \cite{xenophon2018open,zheng2022multi} to circumvent issues related to accessing Critical Energy/Electric Infrastructure Information (CEII).

\subsection*{Adaptive scaling of climate models informed by power systems}

Power system researchers can assist climate researchers in addressing climate change through power grid design and operation in two major ways. First, a power grid operational model (e.g., demand and generation embedded in weather patterns) provides higher-resolution carbon accounting for the energy sector. Monitoring power grid demands provides insights into the emissions from industrial or commercial sectors, where electricity consumption is directly related to emissions. This is because power consumption not only serves as a barometer of societal behavior and prosperity but also underpins economic activities within the industrial and commercial sectors. Climate researchers need to explore how emission changes at a regional scale due to electric grid changes can have an impact on climate \cite{foley2016climate}.

Second, developing a global high-resolution climate model is computationally very expensive \cite{yeager2021bringing,chang2020unprecedented}; however, the spatial resolution of climate models can be effectively shaped by the specific needs of power system tasks. Climate models are successfully used to predict climate variability at seasonal-to-decadal (S2D) timescales and project long-term climate changes at decadal-to-centennial (D2C) timescales \cite{noaa_models}. S2D predictions providing information about natural climate variabilities, such as El Niño or La Niña, and near-term trends are sensitive to initial conditions \cite{dalton2013comparison}. D2C projections aim to understand long-term trends. Therefore, S2D predictions can be used to understand the near-term impact of extreme weather events on power grids, while D2C projections can provide insights into long-term transmission system planning and energy security.

Power system engineers have a major role to play by providing climate scientists with specific regions to downscale their global climate predictions and projections and generate high-resolution climate-power interaction models for decision-making while ensuring reasonable computational efficiency. For example, providing the downscaling of S2D predictions to hurricane-prone coastal city regions would offer much-needed decision support for the planning of both transmission and generation assets in power systems. Similarly, downscaling to the load center and renewable-rich areas from D2C projections could have higher priority from the energy security perspective. Challenges persist due to the limited resolution of current-generation global models, typically around 100 km, which hinders accurate forecasting of extreme weather statistics \cite{schwierz2006challenges,alizadeh2022advances}. This limitation results in significant uncertainties in predictions and projections of changes in extreme weather.

\subsection*{Climate-aware market redesign for power systems}

Many countries around the world went through the deregulation process to introduce competition in providing consumers with reliable and affordable electrical energy. Because of historical reasons, the degree of deregulation varied. For example, the United States contains a mix of vertically integrated regulated monopoly regions, vertically disintegrated energy markets, and energy markets, allowing the participation of traditional utilities without fully committing to deregulation \cite{borenstein2015us}. Regulated, vertically integrated utilities may not follow a cost-minimizing approach \cite{gowrisankaran2024energy}, and consumers may not enjoy the benefits of lower energy costs. Limited oversight on energy-only markets and various capacity markets might not provide resource adequacy for day-to-day power grid operations in deregulated environments. In deregulated regions, grid utilities may not be allowed to own generation (including renewables) largely because of market rules \cite{EPA2024}. Aside from variabilities from renewable energy generation, market operators face major challenges as more and more marginal cost generators are replaced with infra-marginal renewable generators \cite{EPRS2023}.

We highlight three innovative market designs to improve resource adequacy issues in deregulated environments, which could be enabled by improved climate models. Firstly, the capacity market is one way of ensuring that all generators will be present to participate in the spot market. Regarding renewable generation participation in capacity markets, Joskow \cite{joskow2022hierarchies} discussed a hybrid framework for capacity and spot markets, where long-term power purchase agreements with wind, solar, and storage developers are competitively procured, which the author called the ``competition for the market,'' and incorporating it with ``competition in the market'' through short-term energy markets. Access to a good climate model would raise confidence among renewable energy developers to develop more renewable energy farms and participate in the market without needing additional support. Secondly, enabled by technological innovations on the consumer end and improved accuracy of consumption patterns, we could frequently see electricity demand side participation in the real world, e.g., load resources participation in Texas electricity market \cite{9208672}, energy coupon \cite{leeTargetedDemandResponse2022}, and direct load control \cite{yilmaz2022analysing}. There has been an increasing demand for an edge-based market that allows consumers to trade their excess energy and integrate it with the wholesale energy market. Thirdly, as we integrate more and more renewable power plants and continue to electrify other sectors, we may run out of transmission line capacity. Innovative solutions with storage (e.g., deploying storage devices at both ends of the transmission lines) enabled by data analytics and climate models would facilitate optimal utilization of transmission resources. Alternatively, energy efficiency \cite{usdoe_energy_efficiency} is another solution to reduce system-wide demand itself.

Markets enable the efficient exchange of resources, but such an efficient exchange may not lead to adopting climate-friendly technologies. This necessitates implementing a performance-based market design that not only incentivizes short-term efficient and reliable operation of the energy system but is also sustainable and adaptive to long-term environmental goals. Suitable regulatory frameworks and financial incentives that encourage energy providers to invest in renewable energy sources or technologies that improve grid reliability and efficiency need to be set up in this regard. Greater attention should be given to balancing the development of new market mechanisms with the potential consequences of heightened uncertainty and complexity, which requires crafting policies that offer the right incentives to the right participants. Power grid operators have to adapt urgently to prevailing weather and climate patterns to effectively implement the energy transition with reliable supply by stating deadlines in terms of renewables and storage. Moreover, there is a need to allow for the extreme weather events arising from the changing climate, at least until greenhouse gas emission is arrested, by building resilience into the system through more accurate information about changes in extreme weather patterns and statistics arising from the changing climate. Power grid researchers should also consider prudent risk-aware scenarios, where power system planning and operation have to adapt to the extreme climate risk environment, such as massive human migrations to habitable parts of the planet or potential for war due to lack of energy.

\section*{Conclusion}

Tackling climate change requires aggressive and timely decarbonization across the entire economy. Decarbonizing the electricity sector, and electrifying other sectors of energy consumption will play a crucial role in this transition. Research in climate models could better inform the planning of energy sources, demand, and the grid. Conversely, specific needs of the electric energy system could also define new research opportunities in higher-resolution climate modeling and simulation. A significant research partnership between the power systems community and the climate change community at large would help with the mitigation of climate risks through an accelerated decarbonization process, while a whole-of-system approach that encompasses the physical, climate and weather, economic, and social systems to develop a scenario-based approach where decisions will ensure a smooth transition.



\bmhead{Acknowledgments} The work of L.X., S.M., Q.Z. and P.C. is supported in part by Texas A\&M Energy Institute, College of Arts and Sciences at Texas A\&M University, and Texas A\&M Engineering Experiment Station.  The work of T.H. is supported by US National Science Foundation Grant 2328205. The work of L.X. is also supported in part by US National Science Foundation Grant ECCS-2038963.

\bmhead{Author contributions} L.X., S.M. T.H. and Q.Z. conceived and designed the paper. P.C. and D.J.H. contributed material and analysis tools. L.X., S.M. T.H., and Q.Z. drafted the paper with input from all co-authors. L.X., S.M. T.H., Q.Z., P.C., D.J.H., and M.S. read and approved the final version of the paper.

\bmhead{Competing interests} The authors declare no competing interests.

\bmhead{Additional information}  
\textbf{Correspondence} should be addressed to Le Xie.

\bibliographystyle{sn-mathphys}
\bibliography{sn-bibliography.bib}

\end{document}